\documentclass[preprint,tightenlines,floats,aps,amsmath,amssymb,nofootinbib,prd,superscriptaddress]{revtex4}
\usepackage{graphicx} 
\usepackage{subfigure}

\usepackage{multirow}
\usepackage{dcolumn} 
\usepackage{bm}
\usepackage{hyperref}
\usepackage{float}
\usepackage{graphicx}
\usepackage{graphics}
\usepackage{epstopdf}
\usepackage{epstopdf}
\usepackage{amsmath}
\usepackage{amssymb}
\usepackage{float}
\usepackage{enumerate}

\graphicspath{{photo/}}

\begin{document}

\title{Anisotropic pressure effect on central EOS of PSR J0740+6620 in the light of dimensionless TOV equation}

\author{Zhihao Yang}
		\affiliation{School of Physics and Optoelectronics, South China University of Technology, Guangzhou 510641, P.R. China}
		\author{ Dehua Wen\footnote{Corresponding author. wendehua@scut.edu.cn}}
		\affiliation{School of Physics and Optoelectronics, South China University of Technology, Guangzhou 510641, P.R. China}
		\date{\today}

\begin{abstract}

It is generally agreed upon that the pressure inside a neutron star is isotropic. However, a strong magnetic field or superfluidity suggests that the pressure anisotropy may be a more realistic model. We derived the dimensionless TOV equation for anisotropic neutron stars based on two popular models, namely the BL model and the H model, to investigate the effect of anisotropy. Similar to the isotropic case, the maximum mass $M_{max}$ and its corresponding radius $R_{Mmax}$ can also be expressed linearly by a combination of radial central pressure $p_{rc}$ and central energy density $\varepsilon_{c}$, which is insensitive to the equation of state (EOS). We also found that the obtained central EOS would change with different values of $\lambda_{BL}$ ($\lambda_{H}$), which controls the magnitude of the difference between the transverse pressure and the radial pressure. Combining with observational data of PSR J0740+6620 and comparing to the extracted EOS based on isotropic neutron star, it is shown that in the BL model, for $\lambda_{BL}$ = 0.4, the extracted central energy density $\varepsilon_{c}$ changed from 546 -- 1056 MeV/fm$^{3}$ to 510 -- 1005 MeV/fm$^{3}$, and the extracted radial central pressure $p_{rc}$ changed from 87 -- 310 MeV/fm$^{3}$ to 76 -- 271 MeV/fm$^{3}$. For $\lambda_{BL}$ = 2, the extracted $\varepsilon_{c}$ and $p_{rc}$ changed to 412 -- 822 MeV/fm$^{3}$ and 50 -- 165 MeV/fm$^{3}$, respectively. In the H model, for $\lambda_{H}$ = 0.4, the extracted $\varepsilon_{c}$ changed to 626 -- 1164 MeV/fm$^{3}$, and the extracted $p_{rc}$ changed to 104 -- 409 MeV/fm$^{3}$. For $\lambda_{H}$ = 2, the extracted $\varepsilon_{c}$ decreased to 894 -- 995 MeV/fm$^{3}$, and the extracted $p_{rc}$ changed to 220 -- 301 MeV/fm$^{3}$.
\end{abstract}

	\maketitle

\section{Introduction} \label{sec:intro}
Constraining the equation of state (EOS) of neutron stars (NSs) is one of the most fundamental and significant problems in nuclear physics and astrophysics \cite{rev1,rev2,rev3,rev4,rev5}. Over the last few years, observation of several objects such as GW170817 \cite{GW170817}, PSR J0740+6620 \cite{66203,66202,66201,66204} and PSR J0030+0451 \cite{04511,04512} creates excellent opportunities for the study of EOS of NSs.

        It is a common assumption that the pressure inside a NS is isotropic. However, observation of exotic compact objects such as the secondary component of GW190814 with a measured mass between 2.50 and 2.67 $M_{\odot}$ ($M_{\odot}$ is the solar mass) \cite{GW190814}, and the black widow pulsar PSR J0952--0607 with a measured mass of $M$ = 2.35 $\pm$ 0.17 $M_{\odot}$ \cite{0607} have aroused the interest in anisotropic NSs. It is concluded by \cite{LHerrera} that even if the system is initially assumed to be isotropic, physical processes are expected in stellar evolution like dissipative fluxes, energy density inhomogeneities, or the appearance of shear in the fluid flow, will always tend to produce pressure anisotropy. The occurrence of pressure anisotropy due to some exotic phenomena, such as pion-condensation \cite{pion1,pion2} or kaon-condensation \cite{kaon}, strong magnetic field \cite{magnetic2,magnetic9,magnetic10,magnetic1,magnetic5,magnetic7,magnetic11,magnetic4,magnetic3,magnetic12,magnetic6,magnetic8}, the existence of a solid core \cite{solid1,solid2}, a mixture of two fluids \cite{two}, superfluidity \cite{superflu}, and viscosity may be a source of local anisotropy \cite{viscosity1,viscosity2,viscosity3,viscosity4}, etc. Please see Ref. \cite{review1,review2} for a review. There is also much research focused on the anisotropic NSs \cite{res1,res2,res3,res4,res5}.

       Since many properties of NSs depend directly on the choice of the EOS, there is still a great deal of uncertainty in the understanding of the EOS of the high-density nuclear matter, which means that there is also a great deal of uncertainty in the NS properties. To constrain EOS from NS observation, one can use machine learning \cite{ML1,ML2,ML3}, Bayesian Inference \cite{Bay1,Bay2}. Besides, it is expected to establish an EOS-insensitive universal relation between the observables and unobservables. Over the past few decades, many relations have been built, such as the I-LOVE-Q relations \cite{UR1}, quasi-universal relations for static and rapid rotating NSs \cite{UR2}, relations between EOS parameters and canonical NSs \cite{UR3}, and the relations between other properties \cite{UR4,UR5,UR6,UR7,UR8,UR9,UR10,UR11}. In addition, the universal relations between anisotropic NSs are also studied \cite{ANUR1,ANUR2}.

       The TOV equation helps us calculate the NS global properties based on EOS, conversely, using the TOV equation to obtain EOS constraint from NS observation might be worth a try. Recently, the scaled TOV equation has been deduced and then used to constrain the core states and access to the ultimate limit for the pressure and energy density \cite{DTOV1,DTOV2}, which offered us a novel approach to insight into the relations between the global properties of NSs and the core EOS through universal relations, and have a high consistency with previous research in understanding the basic problems about NSs. Considering the existence of anisotropic NSs cannot be excluded by current observation, and previous research focusing on anisotropic NSs only give out how the pressure anisotropy affect the NS global properties but the effect on the extraction of the NS core EOS is not given any further, we try to solve it in this work.

In the present work, we aim to deduce the dimensionless TOV equation for anisotropic NS, and further investigate the anisotropy effect while extracting the central EOS of PSR J0740+6620. 36 EOSs with a maximum mass larger than 1.8 $M$$_{\odot}$ are employed in this work. These include (1) 5 consistent relativistic mean-field (CRMF) EOSs, i.e., G2* \cite{G2s}, IU-FSU \cite{IUFSU}, TW99 \cite{TW99}; (2) 19 DDRMF and NLRMF EOSs such as DD \cite{DD}, DD-ME1\cite{DDME1}, DD-LZ1 \cite{DDLZ1},  NL1 \cite{NL1NL2}, NLZ \cite{NLZNLZ2}, NLSV1 \cite{NLSV1NLSV2}, TM1\cite{TM1}; (3) 9 microscopic EOS such as ALF2 \cite{ALF2} EOS based on hybrid (nuclear+quark) matter, APR3 EOS \cite{APR} based on variational calculations of two-nucleon and three-nucleon interactions, ENG \cite{ENG} and MPA1 \cite{MPA1} EOS based on the Dirac-Brueckner-Hartree-Fock method, WFF1 EOS \cite{WFF} based on different two-nucleon and three-nucleon potentials; (4) 3 EOS based on the quark mean-field bag (QMFB) model by incorporating the bag confinement mechanism, i.e., QMFL40, QMFL60, and QMFL80 \cite{QMFL}.

        This paper is organized as follows. In Sec. \ref{sec:DTOV}, the process of deducing the dimensionless TOV equation of anisotropic NS is presented. In Sec. \ref{sec:Extract}, the effect of anisotropy in extracting the central EOS is given and discussed. Finally, a summary is given in Sec. \ref{sec:summary}. To simplify the equations, the geometric units $\mathit{G}$ = $\mathit{c}$ =1 is adopted.

\section{Dimensionless TOV equation of anisotropic neutron star} \label{sec:DTOV}
	\subsection{TOV equation for anisotropic neutron star}

Considering a static and spherically symmetric equilibrium distribution of matter, the energy-momentum tensor ($T^{\mu\nu}$) is defined as \cite{range1,range2}
\begin{equation}
			\label{eq1}
				T^{\mu\nu}=p_{t}g^{\mu\nu}+(\varepsilon+p_{t})u^{\mu}u^{\nu}+(p_{r}-p_{t})k^{\mu}k^{\nu},
		\end{equation}
where $g^{\mu\nu}$ is the space-time metric, $u^{\mu}$ is the 4-velocity of the fluid, $k^{\mu}$ is the radial unit vector, $p_{t}$, $p_{r}$ and $\varepsilon$ is the tangential pressure, radial pressure and energy density. These 4-vectors satisfy the following conditions
\begin{equation}
			\label{eq2}
   k^{\mu}k_{\mu}=1,
   u^{\mu}k_{\mu}=0.
		\end{equation}

The Schwarzschild metric for the star having a spherically symmetric and static configuration is described as
\begin{equation}
			\label{eq3}
  ds^{2}=e^{\nu}dt^{2}-e^{\lambda}dr^{2}-r^{2}d\theta ^{2}-r^{2}sin^{2}\theta d\varphi^{2}.
		\end{equation}

To exactly calculate the global properties of NS, one should indeed take into account the magnetic field, rotation, and thus deformation that deviates from spherical symmetry. However, it is stimulated by Pattersons et. al that even with strong anisotropy and slow rotation, with the mass increase, the star is getting more spherical\cite{Rec1}. Overall, since we ignore the effect of rotation and focus on the maximum mass of anisotropic neutron star, we assume spherical symmetry is still valid.  Thus the modified Tolman-Oppenheimer-Volkoff (TOV) equation can be obtained by solving Einstein's equations as \cite{BL}
		\begin{equation}
			\label{eq4}
				\frac{dp_{r} }{dr} =-\frac{(\varepsilon +p_{r})(m+4\pi r^{3}p_{r} )}{r^{2}-2mr} +\frac{2 \sigma}{r},
		\end{equation}
  \begin{equation}
			\frac{dm}{dr} =4\pi r^{2}\varepsilon,
			\label{eq5}
		\end{equation}
  where $\sigma$ $\equiv$ $p_{t}$ - $p_{r}$ is the anisotropy parameter, and $\mathit{m}$ is the enclose mass corresponding to radius $\mathit{r}$. Equation (\ref{eq4}) and (\ref{eq5}) can be integrated from center $\mathit{r}$ =0, $\mathit{m}$ = 0, $p_{r}$ = $p_{rc}$ ($p_{rc}$ is the radial central pressure) to the surface $\mathit{r = R}$, $\mathit{m = M}$, $p_{r}$ = 0.

			\subsection{Models of anisotropic neutron star}
  In this work, we use two popular model, namely the BL model \cite{BL}and the H model\cite{H}, these two models are based on the assumptions that\cite{con1,con2}:

  (i) The anisotropy should vanish at the origin, i.e. $P_{r}=P_{t}$ at $r$=0;

  (ii) The pressure and energy density must be positive, i.e. $P_{r}$, $P_{t}$, $\varepsilon$ $<$0;

  (iii) The radial pressure and energy density must be monotonically decreasing, i.e.$\frac{\mathrm{d}P_{r}}{\mathrm{d} r}$ and $\frac{\mathrm{d}\varepsilon }{\mathrm{d} r}$ $<$0;

  (iv)The anisotropic fluid configurations with different conditions such as the null energy ($\varepsilon$ $>$ 0), the dominant energy ($\varepsilon$+$P_{r}$ $>$ 0, $\varepsilon$+$P_{t}$ $>$ 0), and the strong energy ($\varepsilon$+$P_{r}$+2$P_{t}$ $>$ 0) must be satisfied inside the star;

  (v)The speed of sound inside the star must obey $0 <c_{s,r}^{2}< 1$, $0 <c_{s,t}^{2}< 1$, where $c_{s}^{2}$ =$\frac{\partial P}{\partial \varepsilon }$.

For the BL model, the relation between $p_{t}$ and $p_{r}$ is
given as
  \begin{equation}
      \label{eq6}
			p_{t} =p_{r}+\frac{\lambda _{BL} }{3} \frac{(\varepsilon+3p_{r})(\varepsilon+p_{r})r^{3}}{r-2m},
  \end{equation}
   where $\lambda_{BL}$ represent the measure of anisotropy. Previous research has given the limit for the BL model, -2$\le$$\lambda_{BL}$$\le$2 \cite{range1}. With the increased positive value of $\lambda_{BL}$, the model will give out a bigger tangential pressure, resulting in a bigger mass at a fixed radius. However, with the decrease of the negative value of $\lambda_{BL}$, the model will give out a smaller tangential pressure thus a smaller mass with a smaller corresponding radius, and may lead to a negative tangential pressure. In this work, we only consider the positive value of $\lambda_{BL}$.
\begin{figure}[ht!]
\centering
\includegraphics[width=0.8\columnwidth]{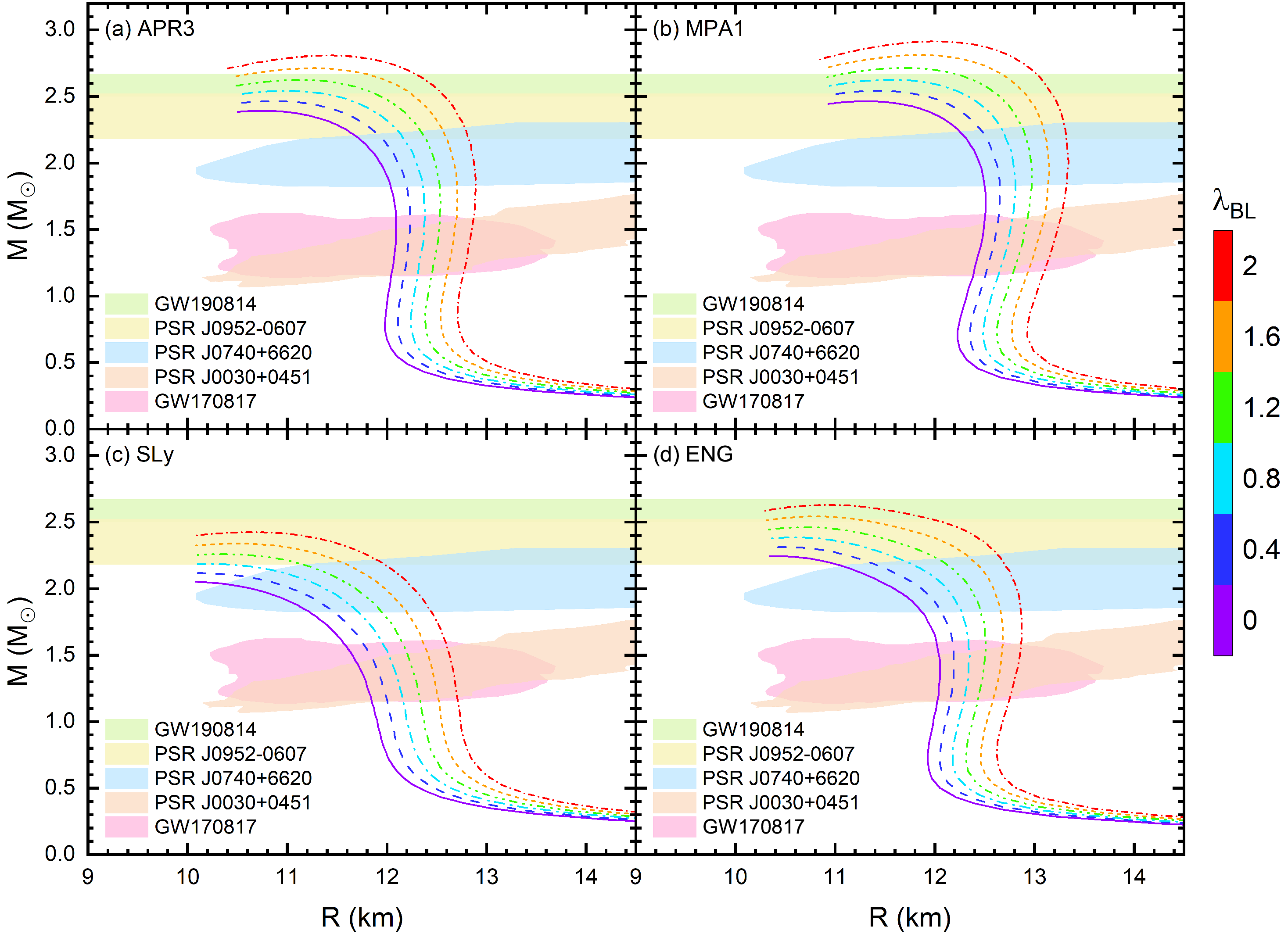}
\caption{$M-R$ curves for anisotropic NSs based on the BL model. Different colours of the curves represent different values of $\lambda_{BL}$, where the red short-dotted one is the highest anisotropic, and the purple solid one is isotropic. The green rectangle is the mass constraint from GW190814 secondary with a mass $M$ = 2.59$_{-0.09}^{+0.08}$ $M_{\odot}$ \cite{GW190814}. The yellow rectangle is the mass constraint from black widow pulsar PSR J0952--0607 with a mass $M$ = 2.35$_{-0.17}^{+0.17}$ $M_{\odot}$ \cite{0607}. The pink, orange, and blue areas correspond to the mass-radius constraint from GW170817 \cite{GW170817}, PSR J0030+0451 \cite{04511}, PSR J0740+6620 \cite{66202}.  }
\label{fig:1}
\end{figure}

To illustrate the effect of anisotropy, the $M-R$ relation of several selected EOSs based on the BL model is shown in Fig. \ref{fig:1}. Each coloured region corresponds to the observational constraint, and different colours of the curves represent different values of $\lambda_{BL}$, where the purple solid one refers to an isotropic NS and the red short-dotted one is the highest anisotropic. It can be seen that the introduction of anisotropy plainly increases the maximum mass and the corresponding radius of the NSs since the positive value of $\lambda_{BL}$ in the BL model means a bigger tangential pressure that can stabilize the structure. Nevertheless, even with such a great change in the $M-R$ curve, all the results still satisfy the observational constraint. Moreover, the introduction of anisotropy makes the $M-R$ curve fall into the interval of GW190814 secondary, which provides a possible mechanism to support the currently observed most massive NS.

  For the H model, ref.\cite{range2} also gives the limit, -2$\le$$\lambda$$_H$$\le$2. For comparison, we only adopt that 0$\le$$\lambda$$_H$$\le$2. The relation between $p_{t}$ and $p_{r}$ is given by
		\begin{equation}
			\label{eq7}
		p_{t} =p_{r}-2\lambda _{H}p_{r}\frac{m}{r}.
		\end{equation}

\begin{figure}[ht!]
  \centering
\includegraphics[width=0.8\columnwidth]{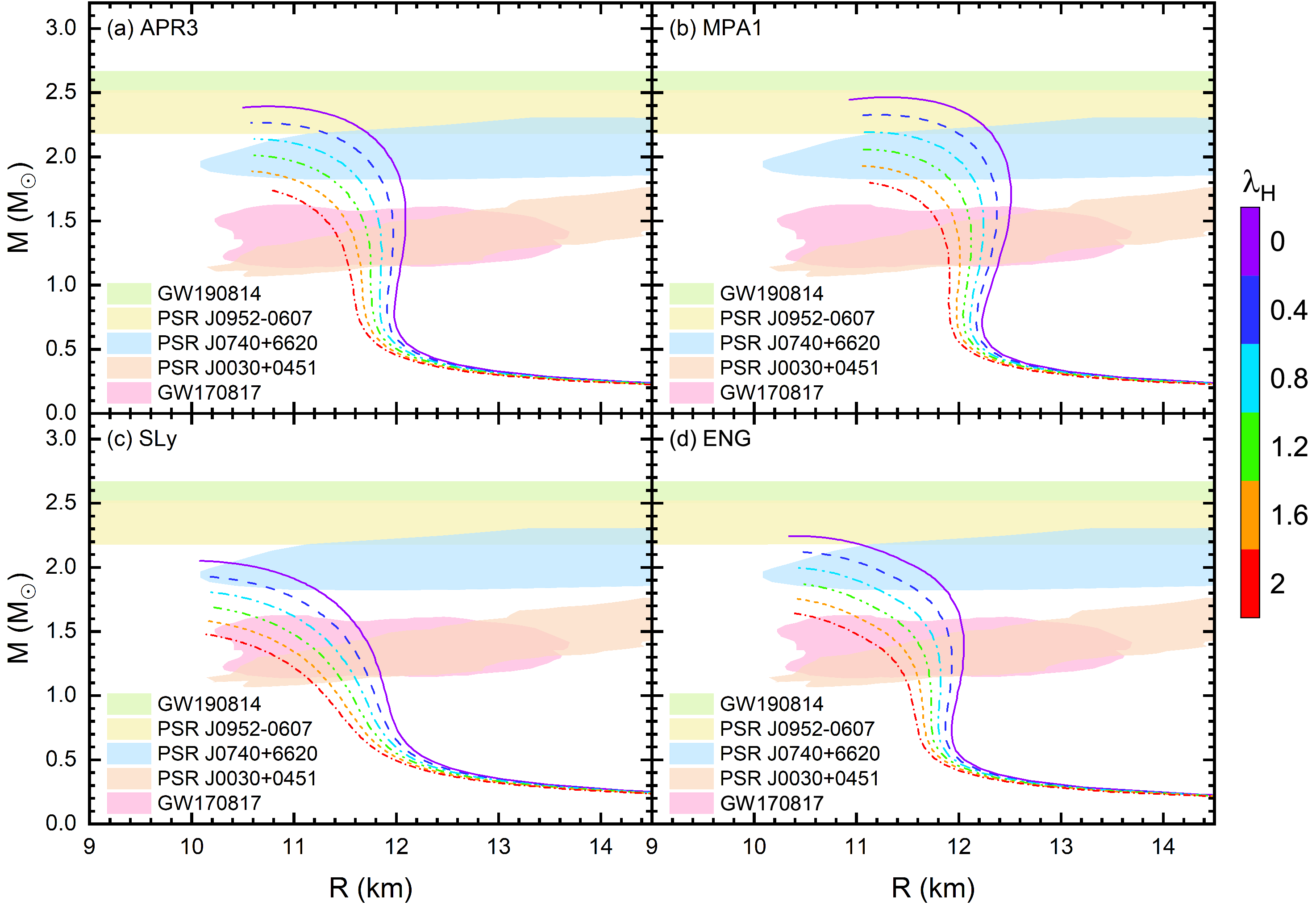}

			\caption{Same as Fig. \ref{fig:1}, but for the H model.}
			\label{fig:2}
		\end{figure}

The $M-R$ relation of the same selected EOSs based on the H model is shown in Fig. \ref{fig:2}. Compared with Fig. \ref{fig:1}, it is shown that the introduction of anisotropy based on the H model decreases the maximum mass and the corresponding radius of NSs, since the positive value of $\lambda _{H}$ means a smaller tangential pressure, resulting in a less massive NS.

\subsection{Dimensionless TOV equation for anisotropic neutron star}

To constrain the central EOS of NSs matter by using certain astrophysical data such as observed NSs radii and masses without using any specific EOS, ref.\cite{DTOV1,DTOV2} propose a new way to insight into NSs core and provides us a direct way to explain the universal relation, which is developed from analyzing perturbatively the dimensionless Tolman-Oppenheimer-Volkoff (TOV) equation. However, considering the effect of the anisotropic pressure mentioned above, some corrections to the dimensionless TOV equation need to be proposed. 		

Noting that ${G}$ = ${c}$ =1, define $S$ = $(4\pi\varepsilon_{c})^{-\frac{1}{2}}$ ($\varepsilon _{c}$ is the central energy density), then one can get the reduced mass $\widehat{m}\equiv{m}/{S}$, radius $\widehat{r}\equiv{r}/{S}$, radial pressure $\widehat{p}_{r}\equiv{p_{r}}/{\varepsilon _{c}}$, tangential pressure $\widehat{p}_{t}\equiv{p_{t}}/{\varepsilon_{c}}$, energy density $\widehat{\varepsilon} \equiv{\varepsilon}/{\varepsilon _{c}}$, which changes the Eq. (\ref{eq4}) and Eq. (\ref{eq5}) as
	
		\begin{equation}
			\label{eq8}
			\frac{d\widehat{p}_{r}}{d\widehat{r}} =-\frac{(\widehat{\varepsilon}+\widehat{p}_{r})(\widehat{m}+\widehat{r}^{3}\widehat{p}_{r})}{\widehat{r}^{2}-2\widehat{m}\widehat{r}} +\frac{2(\widehat{p}_{t}-\widehat{p}_{r})}{\widehat{r}},
		\end{equation}
		\begin{equation}
			\label{eq9}
			\frac{d\widehat{m}}{d\widehat{r}} =\widehat{r}^{2}\widehat{\varepsilon}.
		\end{equation}
		
As there exists a difference in tangential pressure between the different models, the BL model and H model will be discussed separately. For the BL model, according to Eq. (\ref{eq6}), by using the reduced radial pressure $\widehat{p}_{r}$, reduced tangential pressure $\widehat{p}_{t}$ and other properties in reduced form, Eq.(\ref{eq6}) now becomes
        \begin{equation}
			\label{eq10}
			\widehat{p}_{t} =\widehat{p}_{r}+\frac{\lambda _{BL} }{12\pi} \frac{(\widehat{\varepsilon}+3\widehat{p}_{r})(\widehat{\varepsilon}+\widehat{p}_{r})\widehat{r}^{3}}{\widehat{r}-2\widehat{m}}.
		\end{equation}

Since we only focus on the EOS of the core region and the densest matter inside the NS, which is close to the origin, in this case, the radial coordinate can be viewed as a small parameter and thus can be expanded. The $\widehat{\varepsilon}$, $\widehat{p}_{r}$, $\widehat{p}_{t}$, $\widehat{m}$ can be expanded in polynomials in terms of dimensionless radial coordinates,
        \begin{eqnarray}
			\label{eq11}
		\widehat{\varepsilon} &=& 1+a_{1} \widehat{r}+a_{2} \widehat{r}^{2}+a_{3} \widehat{r }^{3}+\cdots, \\
\widehat{p}_{r} &=& \widehat{p}_{rc}+b_{1} \widehat{r}+b_{2} \widehat{r}^{2}+b_{3} \widehat{r }^{3}+\cdots, \\
        \widehat{p}_{t} &=& \widehat{p}_{tc}+c_{1} \widehat{r}+c_{2} \widehat{r}^{2}+c_{3} \widehat{r }^{3}+\cdots,\\
        \widehat{m} &=& d_{1} \widehat{r}+d_{2} \widehat{r}^{2}+d_{3} \widehat{r}^{3}+\cdots.
		\end{eqnarray}
Putting them into the Eq. (\ref{eq8})--Eq. (\ref{eq10}), matching the coefficients, and ignoring the higher order (For we only focus on the core region of the maximum mass of NS, which will be mostly governed by the core region, the ignorance of the higher order will lead to a small effect, as shown in the APPENDIX of \cite{DTOV1}), one has $b_{1}=0, c_{1}=0, d_{1}=0, d_{2}=0, d_{3}=1/3$, and
\begin{equation}
    \label{eq15}
		b_{2} =\frac{1}{6} (\frac{\lambda _{BL}}{2\pi} -1) (\widehat{p}_{rc}+1)(3\widehat{p}_{rc}+1), \; c_{2} =\frac{1}{6} (\frac{\lambda _{BL}}{\pi} -1) (\widehat{p}_{rc}+1)(3\widehat{p}_{rc}+1).
\end{equation}
The boundary condition $\widehat{p}_{r}=0$ means $\widehat{p}_{rc}$+$b_{2}\widehat{r}^{2}$=0,
thus $\widehat{r}=\sqrt{-\widehat{p}_{rc}/b_{2}}$, i.e.,
\begin{equation}
           \label{eq16}		
\widehat{r}=(\frac{6\widehat{p}_{rc}}{(1-\frac{\lambda _{BL}}{2\pi} ) (\widehat{p}_{rc}+1)(3\widehat{p}_{rc}+1)} )^{\frac{1}{2} },
\end{equation}
then multiplied by the scale $S\equiv(4\pi \varepsilon_{c})^{-\frac{1}{2}} \sim \varepsilon_{c}^{-\frac{1}{2}}$, the stellar radius $R \sim \varepsilon_{c}^{-\frac{1}{2}} \widehat{r}$ turn out to be
\begin{equation}
           \label{eq17}		
R\sim \beta \equiv \frac{1}{\sqrt{\varepsilon_{c}}} (\frac{\widehat{p}_{rc}}{(1-\frac{\lambda _{BL}}{2\pi} ) (\widehat{p}_{rc}+1)(3\widehat{p}_{rc}+1)} )^{\frac{1}{2} }.
\end{equation}
Noting that $d_{1}=d_{2}=0, d_{3}=1/3$, thus $\widehat{m}=\widehat{r}^{3}/3$, multiplied by the scale $S$, the stellar mass $M \sim \varepsilon_{c}^{-\frac{1}{2}} \widehat{r}^{3}$ becomes
       \begin{equation}
           \label{eq18}
		M\sim \alpha \equiv \frac{1}{\sqrt{\varepsilon_{c}}} (\frac{\widehat{p}_{rc}}{(1-\frac{\lambda _{BL}}{2\pi} ) (\widehat{p}_{rc}+1)(3\widehat{p}_{rc}+1)} )^{\frac{3}{2}}.
       \end{equation}

To simplify the expression, the symbols $\beta$ and $\alpha$ are used in Eq. (\ref{eq17}) and Eq. (\ref{eq18}) ($\kappa$ and $\gamma$ in Eq. (\ref{eq19}) and Eq. (\ref{eq20})) to denote the formula. For the H model, it has a similar process but a different result, due to their difference in tangential pressure, that is (Please see APPENDIX A for details)
            \begin{equation}
			\label{eq19}
R\sim \kappa \equiv  \frac{1}{\sqrt{\varepsilon_{c}}} (\frac{\widehat{p}_{rc}}{4\lambda _{H}\widehat{p}_{rc}+ (\widehat{p}_{rc}+1)(3\widehat{p}_{rc}+1)} )^{\frac{1}{2}},
		\end{equation}
 \begin{equation}
\label{eq20}
M\sim \gamma \equiv   \frac{1}{\sqrt{\varepsilon_{c}}} (\frac{\widehat{p}_{rc}}{4\lambda _{H}\widehat{p}_{rc}+ (\widehat{p}_{rc}+1)(3\widehat{p}_{rc}+1)} )^{\frac{3}{2}}.
 \end{equation}
   The distinction between Eq. (\ref{eq17}), Eq. (\ref{eq18}) and Eq. (\ref{eq19}), Eq. (\ref{eq20}) lies in the denominator. For the BL model, it adds a coefficient that only depends on the value of $\lambda_{BL}$, while for the H model, it adds a linear term that is related to the value of $\lambda_{H}$ and $\widehat{p}_{rc}$. It is also clear that when $\lambda$$_{BL}$=0 or $\lambda$$_H$=0, Eq. (\ref{eq17})--Eq. (\ref{eq20}) goes back to the isotropic case \cite{DTOV1}.

\begin{figure}[ht!]
\centering
\includegraphics[width=1\columnwidth]{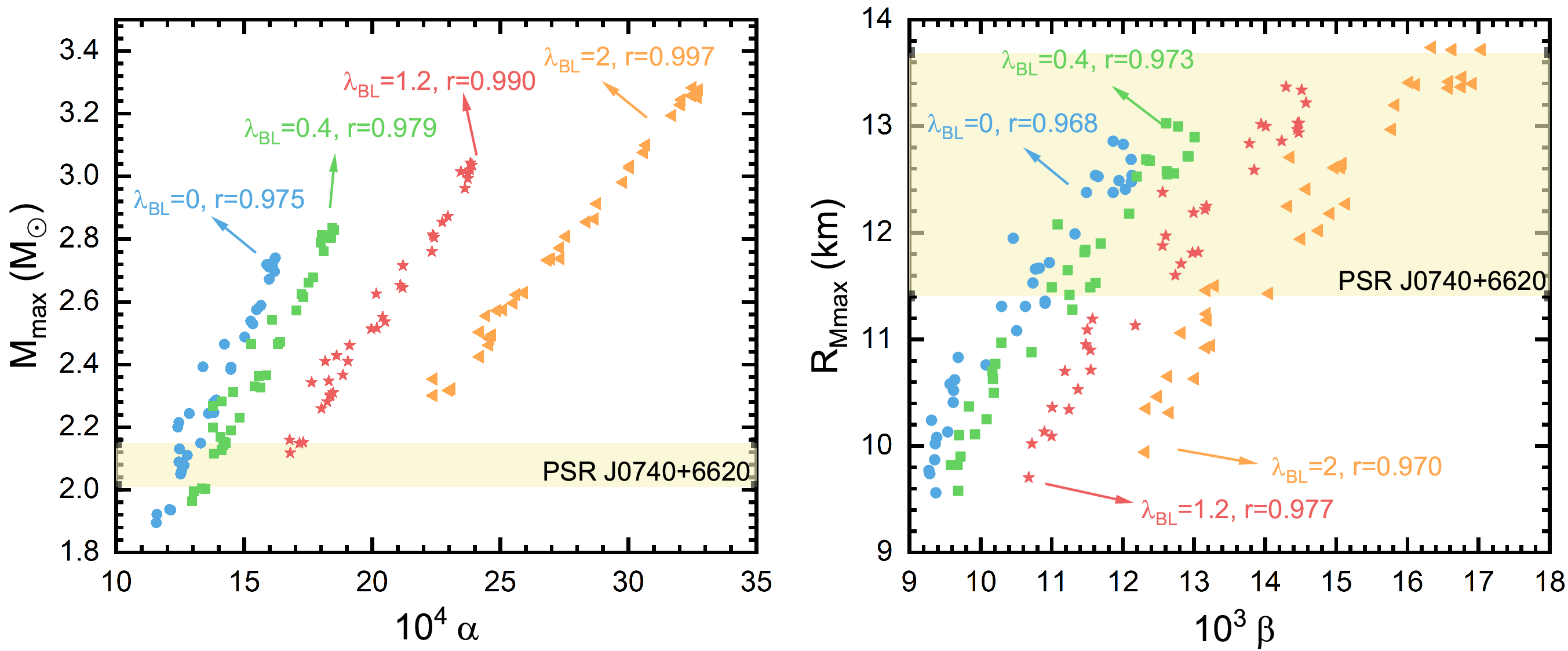}
			\caption{Relations of M$_{max}$ -- $\alpha$ (left) and R$_{Mmax}$ -- $\beta$ (right), calculated by using 36 different EOS introduced above. The milky--yellow region represents the mass and radius constraint from PSR J0740+6620 \cite{66202,66201}, $M$ = 2.08$^{+0.07}_{-0.07}$$M_{\odot}$, $R$ = 12.39$^{+1.30}_{-0.98}$km. }
			\label{fig:3}
		\end{figure}

\begin{figure}[ht!]
			\centering
	\includegraphics[width=1\textwidth]{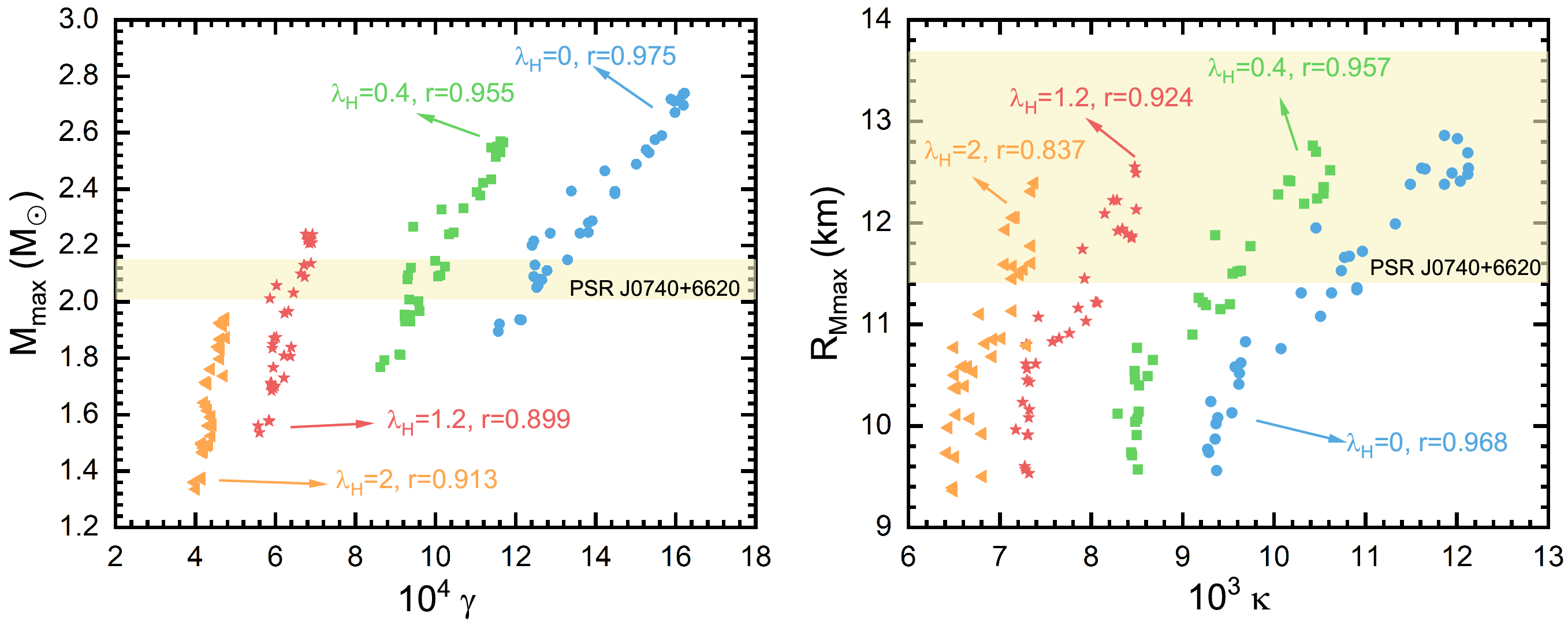}
			\caption{Relation of the M$_{max}$ -- $\gamma$ (left) and R$_{Mmax}$ -- $\kappa$ (right). The meaning of the milky--yellow region is the same as in Fig. \ref{fig:3}.}
			\label{fig:4}
		\end{figure}

The relaions of $M$ -- $\alpha$ ($M$ -- $\gamma$) and $R$ -- $\beta$ ($R$ -- $\kappa$) for a given value of $\lambda_{BL}$ ($\lambda_{H}$) are shown in Fig. \ref{fig:3} (Fig. \ref{fig:4}). Each point represents the maximum mass configuration of the chosen EOS, and each colour of the points corresponds to a particular value of $\lambda_{BL}$ ($\lambda_{H}$), resulting in different degrees of anisotropy. The milky--yellow region represents the mass and radius constraint from PSR J0740+6620, $M$ = 2.08$^{+0.07}_{-0.07}$ $M_{\odot}$, $R$ = 12.39$^{+1.30}_{-0.98}$ km \cite{66201,66202}. For the BL model, the relations of M$_{max}$ -- $\alpha$ and R$_{Mmax}$ -- $\beta$ are shown in Fig. \ref{fig:3}, and its fitting formula for $\lambda$$_{BL}$ = 0, 0.4, 1.2 and 2 are given by
\begin{eqnarray}
    \label{eq21}
M_{\lambda_{BL}=0}^{max} &=& 0.172_{-0.007}^{+0.007}\times10^{4}\alpha-0.062_{-0.094}^{+0.094}, \nonumber\\
		M_{\lambda_{BL}=0.4}^{max} &=& 0.151_{-0.005}^{+0.005}\times10^{4}\alpha+0.040_{-0.084}^{+0.084}, \nonumber\\
M_{\lambda_{BL}=1.2}^{max} &=& 0.123_{-0.003}^{+0.003}\times10^{4}\alpha+0.065_{-0.062}^{+0.062}, \\
        M_{\lambda_{BL}=2}^{max} &=& 0.096_{-0.001}^{+0.001}\times10^{4}\alpha+0.158_{-0.038}^{+0.038}. \nonumber
\end{eqnarray}

\begin{eqnarray}
    \label{eq22}
R_{\lambda_{BL}=0}^{Mmax} &=& 0.969_{-0.043}^{+0.043}\times10^{3}\beta+1.052_{-0.462}^{+0.462}, \nonumber\\
		R_{\lambda_{BL}=0.4}^{Mmax} &=& 0.903_{-0.037}^{+0.037}\times10^{3}\beta+1.323_{-0.414}^{+0.414}, \nonumber\\
R_{\lambda_{BL}=1.2}^{Mmax} &=& 0.826_{-0.031}^{+0.031}\times10^{3}\beta+1.280_{-0.400}^{+0.400}, \\
        R_{\lambda_{BL}=2}^{Mmax} &=& 0.730_{-0.031}^{+0.031}\times10^{3}\beta+1.463_{-0.458}^{+0.458}. \nonumber
\end{eqnarray}

Obviously, for a given value of $\lambda_{BL}$, there exist a linear correlation between M$_{max}$ and $\alpha$ or R$_{Mmax}$ and $\beta$. For M$_{max}$ -- $\alpha$, when $\lambda$$_{BL}$ = 0, 0.4, 1.2, and 2, the Pearson's coefficient is 0.975, 0.979, 0.990, and 0.997, respectively. And for the R$_{Mmax}$ -- $\beta$, when $\lambda$$_{BL}$ = 0, 0.4, 1.2, and 2, the Pearson's coefficient is 0.968, 0.973, 0.977, 0.970. The reason why mass is more dependent on $\alpha$ is that mass is mostly contributed by the core region, while for radius it will be more influenced by the crustal region that does not appear in the expression of $\beta$. It is also important to note that with the increase of anisotropy, the slope of both relations decreases, implying the tendency for mass (radius) to increase with $\alpha$ ($\beta$) slows down. that is owing to the tangential pressure, which plays a more important role in stabilizing the structure.

The relations of the M$_{max}$ -- $\gamma$ and R$_{Mmax}$ -- $\kappa$ for the H model are presented in Fig. \ref{fig:4}, and its fitting formula are given by
 \begin{eqnarray}
    \label{eq23}
M_{\lambda_{H}=0}^{max} &=& 0.172_{-0.007}^{+0.007}\times10^{4}\alpha-0.062_{-0.094}^{+0.094}, \nonumber\\
		M_{\lambda_{H}=0.4}^{max} &=& 0.248_{-0.013}^{+0.013}\times10^{4}\gamma-0.332_{-0.135}^{+0.135}, \nonumber\\
M_{\lambda_{H}=1.2}^{max} &=& 0.478_{-0.040}^{+0.040}\times10^{4}\gamma-1.086_{-0.251}^{+0.251}, \\
        M_{\lambda_{H}=2.0}^{max} &=& 0.779_{-0.060}^{+0.060}\times10^{4}\gamma-1.773_{-0.264}^{+0.264}. \nonumber
\end{eqnarray}

\begin{eqnarray}
    \label{eq24}
R_{\lambda_{H}=0}^{Mmax} &=& 0.969_{-0.043}^{+0.043}\times10^{3}\beta+1.052_{-0.462}^{+0.462}, \nonumber\\
		R_{\lambda_{H}=0.4}^{Mmax}  &=& 1.159_{-0.060}^{+0.060}\times10^{3}\kappa+0.388_{-0.566}^{+0.566}, \nonumber \\
R_{\lambda_{H}=1.2}^{Mmax} &=& 1.741_{-0.123}^{+0.123}\times10^{3}\kappa-2.490_{-0.960}^{+0.960}, \\
        R_{\lambda_{H}=2.0}^{Mmax} &=& 2.213_{-0.249}^{+0.249}\times10^{3}\kappa-4.370_{-1.708}^{+1.708}.  \nonumber
\end{eqnarray}

The meaning of the milky--yellow region is the same as in Fig. \ref{fig:3}. Although the slope changes with the given value of $\lambda _{H}$ again, it is contrary to the trend of the BL model. What's more, the Pearson's coefficient for M$_{max}$ -- $\gamma$ relations when $\lambda$$_{H}$ = 0, 0.4, 1.2 and 2 is 0.975, 0.955, 0.899, 0.913; while 0.968, 0.957, 0.924, 0.837 for the R$_{Mmax}$ -- $\kappa$ relations, respectively. In the BL model, an increasing positive parameter results in an increasing repulsive force that can support a larger mass and is more affected by the core region. However, in the H model, it indicates an attractive force that results in a smaller maximum mass. As a result, the radius will be less affected by the core region, leading to a decreasing Pearson's coefficient for R$_{Mmax}$ -- $\kappa$ relations with increasing $\lambda$$_{H}$.  

\section{Extracting the central equation of state} \label{sec:Extract}
From Eq. (\ref{eq17}) - Eq. (\ref{eq20}), it is shown that the combination of central pressure and energy density can explain the maximum mass and the corresponding radius. Thus the observation of maximum mass and its radius of NS will lead to a constraint on the central pressure and energy density. With further observation of NSs, one can decrease the uncertainty of core-state EOS, which still can not be achieved by ab initio calculation. For the BL model, combining the Eq. (\ref{eq17}), Eq. (\ref{eq18}), Eq. (\ref{eq21}), and Eq. (\ref{eq22}) one can obtain Eq. (\ref{eq25}). One would find that in both sides of the Eq. (\ref{eq25}) it still contains the pressure term, to separate the pressure term, one should guess the functional form between pressure and energy density. 
\begin{equation}
			\label{eq25}
			p_{rc}^{M-constraint}=DA^{\frac{2}{3}}\varepsilon^{\frac{4}{3}}_{c}(3\widehat{p}^{2}_{rc}+4\widehat{p}_{rc}+1), \;
   p_{rc}^{R-constraint}=DB^{2}\varepsilon_{c}^{2}(3\widehat{p}^{2}_{rc}+4\widehat{p}_{rc}+1),
		\end{equation}
and
\begin{equation}
			\label{eq26}
			D\equiv(1-\frac{\lambda _{BL} }{2\pi}), \;
                A\equiv\frac{\frac{M_{max}}{M_{\odot}}-b}{k}, \; B\equiv\frac{\frac{R_{Mmax} }{km}-b}{k},
		\end{equation}
where $k$ and $b$ are the slope and intercept with the y-axis corresponding to each formula in Eq. (\ref{eq21}) and Eq. (\ref{eq22}). Assuming that $p_{rc}$ can also be written in the polynomials of central energy density $\varepsilon_{c}$,
  \begin{equation}
			\label{eq27}
p_{rc}^{M-constraint}=DA^{\frac{2}{3}}\varepsilon_{c}^{\frac{4}{3}}(1+aA^{\frac{2}{3}}\varepsilon_{c}^{\frac{1}{3}}+bA^{\frac{4}{3}}\varepsilon^{\frac{2}{3}_{c}}+cA^{\frac{6}{3}}\varepsilon^{\frac{3}{3}}_{c}+\cdots),
\end{equation}
  \begin{equation}
			\label{eq28}
p_{rc}^{R-constraint}=DB^{2}\varepsilon_{c}^{2}(1+aB^{2}\varepsilon_{c}+bB^{4}\varepsilon^{2}_{c}+cB^{6}\varepsilon^{3}_{c}+\cdots).
		\end{equation}
Putting the Eq. (\ref{eq27}) and Eq. (\ref{eq28}) back into the Eq. (\ref{eq25}) separately, matching their coefficients, the coefficients turn out to be
  \begin{equation}
			\label{eq29}
			a=4D,\; b=19D^{2},\; c=100D^{3},\; \cdots
		\end{equation}
So the Eq. (\ref{eq25}) becomes
\begin{equation}
    \label{eq30}
p_{rc}^{M-constraint} = DA^{\frac{2}{3}}\varepsilon_{c}^{\frac{4}{3}}(1+4DA^{\frac{2}{3}}\varepsilon_{c} ^{\frac{1}{3}}+19D^{2}A^{\frac{4}{3}}\varepsilon_{c}^{\frac{2}{3}}+100D^{3}A^{\frac{6}{3}}\varepsilon_{c} ^{\frac{3}{3}}+\cdots ),
\end{equation}
\begin{equation}
\label{eq31}
p_{rc}^{R-constraint}=DB^{2}\varepsilon_{c}^{2}(1+4DB^{2}\varepsilon_{c}+19D^{2}B^{4}\varepsilon_{c}^{2}+100D^{3}B^{6}\varepsilon_{c}^{3}+\cdots ).
\end{equation}
However, causality should also be considered when constraining the central EOS. Using Eq. (\ref{eq18}) and the boundary condition ${\mathrm{d}M}/{\mathrm{d} \varepsilon _{c} }=0$, the sound speed square is obtained as (see APPENDIX B for details)
\begin{equation}
			\label{eq32}
			c_{s}^2 \equiv \frac{\mathrm{d}p_{rc}}{\mathrm{d} \varepsilon_{c}} =\widehat{p}_{rc}(\frac{1+3\widehat{p}_{rc}^2+4\widehat{p}_{rc}}{3(1-3\widehat{p}_{rc}^2)}+1 ),
		\end{equation}
	which have the same form as the isotropic NS, so that their sound speed constraint does not change with the value of $\lambda _{BL}$, and so have the same limit, $\widehat{p}_{rc}$ = 0.374 \cite{DTOV1}. Using Eq. (\ref{eq30}) -- Eq. (\ref{eq32}), we can now constrain the central EOS.

\begin{figure}
\centering
\includegraphics[width=1\columnwidth]{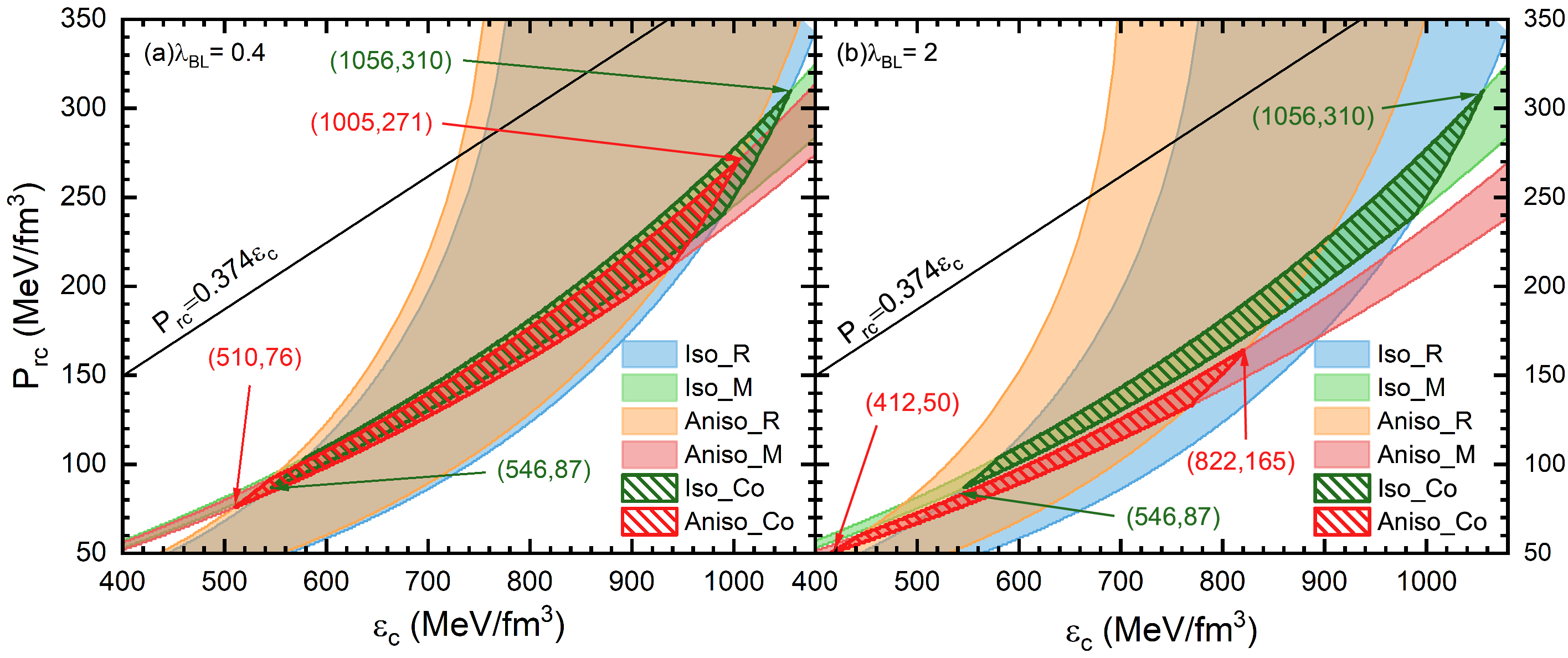}
			\caption{Central EOS of PSR J0740+6620 \cite{66202,66201} extracted from the BL model based on Eq. (\ref{eq30}) and Eq. (\ref{eq31}), for $\lambda _{BL}$ = 0.4 (left) and $\lambda _{BL}$ = 2 (right), compared with the isotropic case. Suffix ``M'' (``R'') represents the result combining mass (radius) observation data with Eq. (\ref{eq30}) (Eq. (\ref{eq31})). The abbreviation ``Iso'' stands for isotropic NS, while the ``Aniso'' stands for anisotropic NS, and the ``Co'' stands for the intersection of mass and radius constraints bands. The black line indicates the causality constraint from Eq. (\ref{eq32}). The green-striped region results from isotropic NS, and the red-striped region from anisotropic NS.}
			\label{fig:5}
		\end{figure}

To show the effect of anisotropy on the extracted central EOS of PSR J0740+6620, the results for the BL model according to Eq. (\ref{eq30}) and Eq. (\ref{eq31}) are shown in Fig. \ref{fig:5}. For $\lambda _{BL}$ = 0.4 in the left panel and $\lambda _{BL}$ = 2 in the right panel, compared to the isotropic case, the striped area indicates the intersection of the $M$ and $R$ constraint. As anisotropy increases, the $R$ constraint gives a stiffer result, while the M constraint yields a softer result. Consequently, the central pressure is smaller for the same energy density, owing to the tangential pressure playing a more significant role in stabilizing the star. For $\lambda_{BL}$ = 0.4, the left panel shows that the extracted central energy density range changed from 546 -- 1056 MeV/fm$^{3}$ to 510 -- 1005 MeV/fm$^{3}$, and the extracted radial central pressure range changed from 87 -- 310 MeV/fm$^{3}$ to 76 -- 271 MeV/fm$^{3}$.
 For $\lambda_{BL}$ = 2, the extracted central energy density changed to 412 -- 822 MeV/fm$^{3}$, and the extracted radial central pressure changed to 50 -- 165 MeV/fm$^{3}$.
 Note that both results are also consistent with the causality constraint.

For the H model, according to Eq. (\ref{eq19}), Eq. (\ref{eq20}), Eq. (\ref{eq23}) and Eq. (\ref{eq24}), the relations change into the following form,
\begin{eqnarray}
    \label{eq33}
		P_{rc}^{M-constraint} =A^{\frac{2}{3}}\varepsilon_{c} ^{\frac{4}{3}}(1+DA^{\frac{2}{3}}\varepsilon_{c} ^{\frac{1}{3}}+(D^{2}+3)A^{\frac{4}{3}}\varepsilon_{c}^{\frac{2}{3}}+(D^{3}+9D)A^{\frac{6}{3}}\varepsilon_{c} ^{\frac{3}{3}}+\cdots ), \\
P_{rc}^{R-constraint}=B^{2}\varepsilon_{c}^{2}(1+DB^{2}\varepsilon_{c}+(D^{2}+3)B^{4}\varepsilon_{c}^{2}+(D^{3}+9D)B^{6}\varepsilon_{c}^{3}+\cdots),
\end{eqnarray}
where
\begin{equation}
			\label{eq35}
			D\equiv4+4\lambda _{H}, \;
                A\equiv\frac{\frac{M_{max}}{M_{\odot}}-b}{k}, \; B\equiv\frac{\frac{R_{Mmax} }{km}-b}{k},
		\end{equation}
and the meanings of $k$ and $b$ are the same as in the BL model.

\begin{figure}
\centering
\includegraphics[width=1\columnwidth]{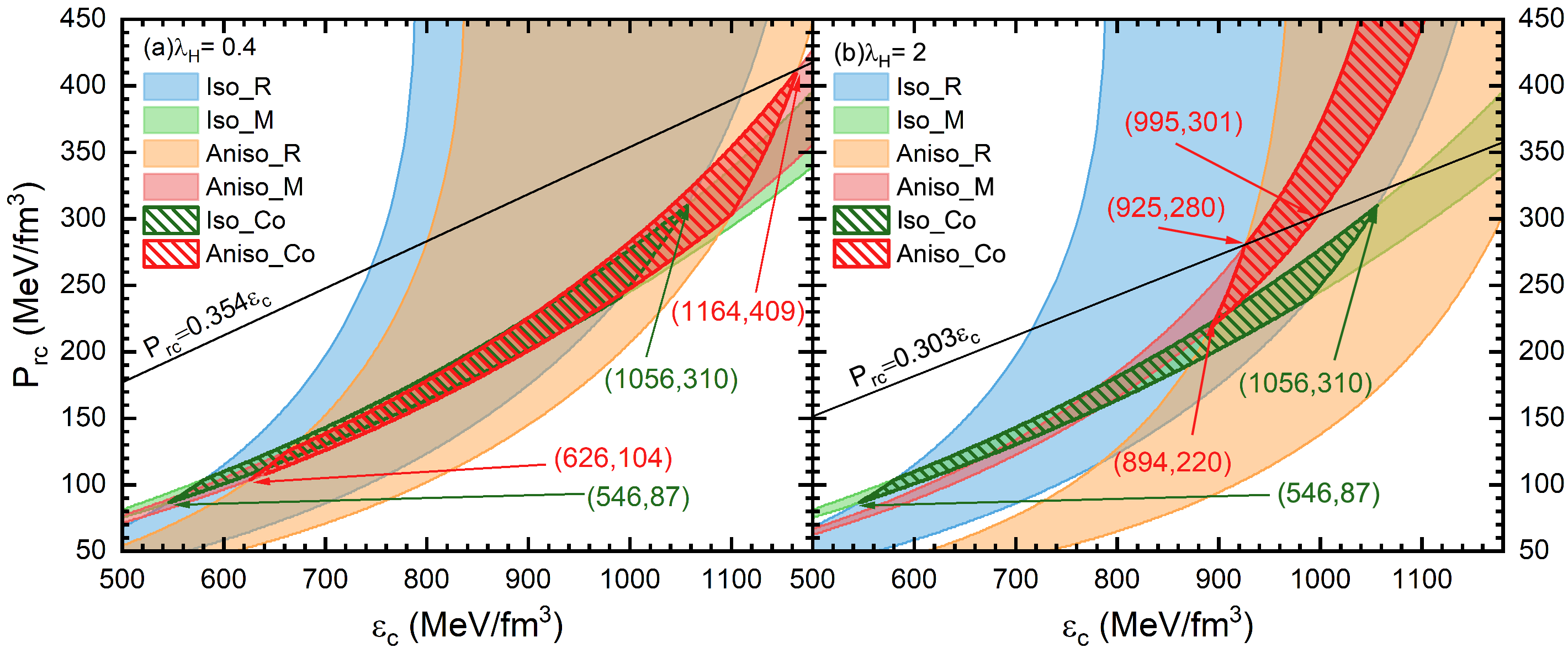}
			\caption{Same as Fig. \ref{fig:5}, but for H model. The causality constraint is from Eq. (\ref{eq36}).}
			\label{fig:6}
		\end{figure}
In addition, the sound speed square of the H model gets some change,
\begin{equation}
			\label{eq36}
			c_{s}^2 \equiv \frac{\mathrm{d}p_{rc}}{\mathrm{d} \varepsilon_{c}} =\widehat{p}_{rc}(\frac{1+3\widehat{p}_{rc}^2+(4+4\lambda _{H} )\widehat{p}_{rc}}{3(1-3\widehat{p}_{rc}^2)}+1).
		\end{equation}

Comparing Eq. (\ref{eq32}) with Eq. (\ref{eq36}), it is found that the two models have different results by causality constraints, for the BL model it does not change with the value of $\lambda_{BL}$, while for the H model it will change with the value of $\lambda_{H}$, thus resulting in different maximum $\widehat{p}_{rc}$. The difference origin at Eq. (\ref{eq6}) and Eq. (\ref{eq7}), then occur in Eq. (\ref{eq17}), Eq. (\ref{eq18}), and Eq. (\ref{eq19}), Eq. (\ref{eq20}), representing different ways about the anisotropy affecting the central radial pressure. The extracted EOS based on the H model is shown in Fig. \ref{fig:6}. As the anisotropy changes, the causality boundary gives different constraints, for $\lambda _{H}$ = 0.4 is $\widehat{p}_{rc}$ = 0.354, but $\widehat{p}_{rc}$ = 0.303 for $\lambda _{H}$ = 2. Also, the $M$ and $R$ constraint bands change with the introduction of anisotropy. For $\lambda_{H}$ = 0.4, the left panel shows that the extracted central energy density range changed from 546 -- 1056 MeV/fm$^{3}$ to 626 -- 1164 MeV/fm$^{3}$, and the extracted radial central pressure range changed from 87 -- 310 MeV/fm$^{3}$ to 104 -- 409 MeV/fm$^{3}$. 
 For $\lambda_{H}$ = 2, the extracted central energy density range changed to 894 -- 995 MeV/fm$^{3}$, and the extracted radial central pressure range changed to 220 -- 301 MeV/fm$^{3}$. 

\begin{figure}[h]
\includegraphics[width=0.5\columnwidth]{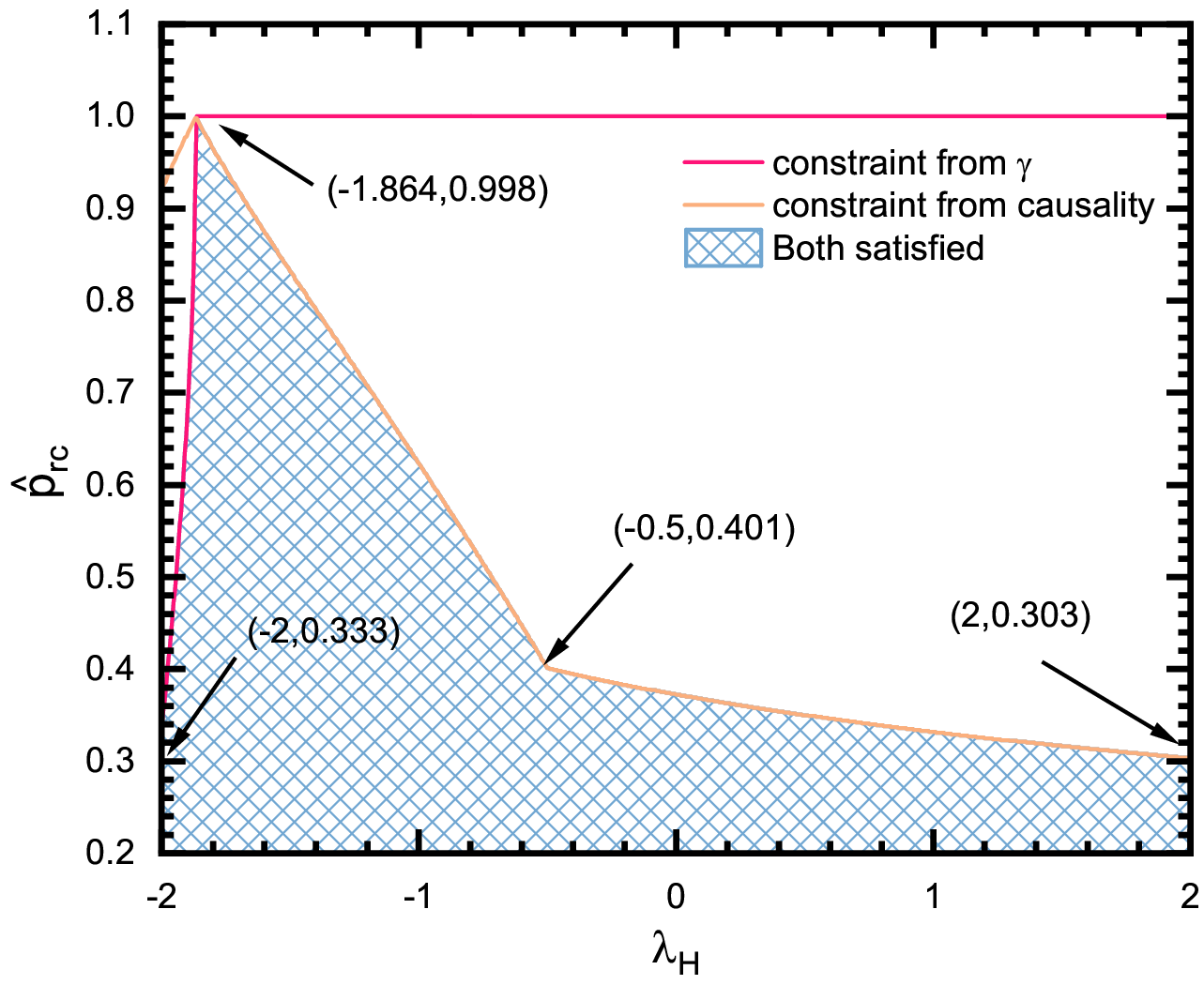}
\centering
			\caption{Maximum $\widehat{p}_{rc}$ as a function of $\lambda _{H}$. The pink line means the constraint of $\gamma$, while the orange one means that causality is obeyed. The blue--striped region means that both conditions are satisfied.}
			\label{fig:7}
		\end{figure}

To directly demonstrate how $\widehat{p}_{rc}$ changes with a given value of $\lambda _{H}$, we deliver Fig. \ref{fig:7}. The pink line means the constraint of $\gamma$, while the orange line means that causality is obeyed. Note that the denominator of Eq. (\ref{eq19}) and Eq. (\ref{eq20}), namely $1+3\widehat{p}_{rc}^2+(4+4\lambda _{H})\widehat{p}_{rc}$, should larger than 0 for each given value of $\lambda _{H}$, and the causality means that Eq. (\ref{eq36}) should larger than 0 and smaller than 1. This gives another constraint that varies with the given $\lambda _{H}$, when both conditions are satisfied it becomes the blue--striped region in Fig. \ref{fig:7}. Therefore, it is noticed that with the changing $\lambda _{H}$, the maximum $\widehat{p}_{rc}$ will change a lot.

\section{summary} \label{sec:summary}

The presence of a strong magnetic field, superfluidity, and other situations will cause anisotropy, leading to an apparent change in the stellar structure. Observations of several exotic compact objects have been made through observation, these fascinating objects possess unknown properties that have motivated the research of anisotropic NSs.

Recently, ref.\cite{DTOV1,DTOV2} derived the dimensionless TOV equation for isotropic NSs, then extracted the central EOS under the observational data of PSR J0740+6620, which has reduced a great uncertainty of the EOS of NSs core matter. In this work, we present the anisotropic dimensionless TOV equation to facilitate the extraction of the central EOS of anisotropic NSs. The result indicates that anisotropy will cause a non-negligible impact on the extraction of the central EOS of PSR J0740+6620. In the BL model, for $\lambda_{BL}$ = 0.4, the extracted central energy density $\varepsilon_{c}$ changed from 546 -- 1056 MeV/fm$^{3}$ to 510 -- 1005 MeV/fm$^{3}$, and the extracted radial central pressure $p_{rc}$ changed from 87 -- 310 MeV/fm$^{3}$ to 76 -- 271 MeV/fm$^{3}$. For $\lambda_{BL}$ = 2, the extracted $\varepsilon_{c}$ and $p_{rc}$ changed to 412 -- 822 MeV/fm$^{3}$ and 50 -- 165 MeV/fm$^{3}$, respectively. In the H model, for $\lambda_{H}$ = 0.4, the extracted $\varepsilon_{c}$ changed to 626 -- 1164 MeV/fm$^{3}$, and the extracted $p_{rc}$ changed 104 -- 409 MeV/fm$^{3}$. For $\lambda_{H}$ = 2, the extracted $\varepsilon_{c}$ changed to 894 -- 995 MeV/fm$^{3}$, and the extracted $p_{rc}$ changed to 220 -- 301 MeV/fm$^{3}$. Additionally, the introduction of anisotropy will result in different causality constraints, for the BL model, it shares the same limit with the isotropic one, $\widehat{p}_{rc}$ = 0.374 \cite{DTOV1}, while for the H model, it changes with the given value of $\lambda_H$, for $\lambda _{H}$ = 0.4 is $\widehat{p}_{rc}$ = 0.354, but $\widehat{p}_{rc}$ = 0.303 for $\lambda _{H}$ = 2. It is worth noting that with the different values of $\lambda_{BL}$ ($\lambda_{H}$), the result will differ a lot. In short, a positive $\lambda_{BL}$ or a negative $\lambda_{H}$ will give out a softer EOS region, while a negative $\lambda_{BL}$ or a positive $\lambda_{H}$) will give out a stiffer EOS region.

Although the existence of a strong magnetic field is considered to be the reason for anisotropy, not all NSs are capable of such strong magnetic fields. Ref.\cite{res6} claim that for quark star, when the baryon density is about 3$n_0$ ($n_0$= 0.16 fm$^{-3}$ is the normal nuclear matter density), the transverse pressure $P_{\bot}$ (which is perpendicular to the magnetic field) and the longitudinal pressure $P_{//}$ (which is parallel to the magnetic field) starts to split out at B$\approx$ 1.5$\times$10$^{17}$ G, B$\approx$4.5$\times$10$^{17}$ G for 5$n_0$ and B$\approx$6$\times$10$^{17}$ G for 7$n_0$. It is also calculated by Ref.\cite{res7} that if one takes the assumption $p_t$=$p_{r}$+$\frac{B^{2}}{4\pi}$, when the central magnetic field is B$_0$$\approx$ 10$^{18}$ G, then $p_t$-$p_{r}$$\le200$MeV$\cdot$fm$^{-3}$. In ref.\cite{res3}, the anisotropy parameter $\lambda_{BL}$ is still independent of the distributions of a magnetic field in the modified BL model. In the subsequent work, it might be worth connecting them and further investigating the effect of magnetic field and anisotropy.

\section{acknowledgement}
This work is supported by NSFC (Grants No. 12375144, 11975101) and Guangdong Natural Science Foundation (Grants No. 2022A1515011552, 2020A151501820).

\appendix \label{appendix}

\section{Derivation of central EOS via polynomial expansion}
One can expand the $\widehat{\varepsilon}$, $\widehat{p_{r}}$, $\widehat{p_{t}}$, $\widehat{m}$ as

\begin{eqnarray}
			\label{eqA1}
		\widehat{\varepsilon} &=& 1+a_{1} \widehat{r}+a_{2} \widehat{r}^{2}+a_{3} \widehat{r }^{3}+\cdots, \\
\widehat{p}_{r} &=& \widehat{p}_{rc}+b_{1} \widehat{r}+b_{2} \widehat{r}^{2}+b_{3} \widehat{r }^{3}+\cdots, \\
        \widehat{p}_{t} &=& \widehat{p}_{tc}+c_{1} \widehat{r}+c_{2} \widehat{r}^{2}+c_{3} \widehat{r }^{3}+\cdots,\\
        \widehat{m} &=& d_{1} \widehat{r}+d_{2} \widehat{r}^{2}+d_{3} \widehat{r }^{3}+\cdots.
		\end{eqnarray}
For the BL model, matching their coefficients according to the Eq. (\ref{eq8})--Eq. (\ref{eq10}), it has $b_{1}=0, c_{1}=0, d_{1}=0, d_{2}=0, d_{3}=1/3$, and
\begin{equation}
			\label{eqA5}
			2b_{2}=-(\widehat{p}_{rc}+1)(\widehat{p}_{rc}+\frac{1}{3})+2(c_{2}-b_{2}),
		\end{equation}
\begin{equation}
			\label{eqA6}
			c_{2}-b_{2}=\frac{\lambda_{BL}}{4\pi} (\widehat{p}_{rc}+1)(\widehat{p}_{rc}+\frac{1}{3}).
		\end{equation}
Thus one can obtain
\begin{equation}
    \label{eqA7}
		b_{2} =\frac{1}{6} (\frac{\lambda _{BL}}{2\pi} -1) (\widehat{p}_{rc}+1)(3\widehat{p}_{rc}+1),\quad c_{2} =\frac{1}{6} (\frac{\lambda _{BL}}{\pi} -1) (\widehat{p}_{rc}+1)(3\widehat{p}_{rc}+1).
\end{equation}
Boundary condition $\widehat{p}_{r}=0$ means $\widehat{p}_{rc}$+$b_{2}\widehat{r}^{2}$=0,
thus $\widehat{r}=\sqrt{-\widehat{p}_{rc}/b_{2}}$, i.e.,
\begin{equation}
           \label{eqA8}		
\widehat{r}=(\frac{6\widehat{p}_{rc}}{(1-\frac{\lambda _{BL}}{2\pi} ) (\widehat{p}_{rc}+1)(3\widehat{p}_{rc}+1)} )^{\frac{1}{2} },
\end{equation}
then multiplied by the scale $S\equiv(4\pi \varepsilon_{c})^{-\frac{1}{2}} \sim \varepsilon_{c}^{-\frac{1}{2}}$, the stellar radius $R$ turn out to be
\begin{equation}
           \label{eqA9}		
R\sim \frac{1}{\sqrt{\varepsilon_{c}}} (\frac{\widehat{p}_{rc}}{(1-\frac{\lambda _{BL}}{2\pi} ) (\widehat{p}_{rc}+1)(3\widehat{p}_{rc}+1)} )^{\frac{1}{2} }.
\end{equation}
Noting that $d_{3}=1/3$, thus $\widehat{m}=\widehat{r}^{3}/3$, multiplied by the scale $S$, the stellar mass $M$ becomes
       \begin{equation}
           \label{eqA10}
		M\sim \frac{1}{\sqrt{\varepsilon_{c}}} (\frac{\widehat{p}_{rc}}{(1-\frac{\lambda _{BL}}{2\pi} ) (\widehat{p}_{rc}+1)(3\widehat{p}_{rc}+1)} )^{\frac{3}{2}}.
       \end{equation}
Once the relation of Eq. (\ref{eqA9}) or Eq. (\ref{eqA10}) is obtained (as we have shown in Eq. (\ref{eq21})-Eq. (\ref{eq22})), it can be changed as
\begin{equation}
           \label{eqA11}
p_{rc}^{M-constraint}=DA^{\frac{2}{3}}\varepsilon^{\frac{4}{3}}_{c}(3\widehat{p}^{2}_{rc}+4\widehat{p}_{rc}+1), \;
   p_{rc}^{R-constraint}=DB^{2}\varepsilon_{c}^{2}(3\widehat{p}^{2}_{rc}+4\widehat{p}_{rc}+1).
\end{equation}
where
\begin{equation}
			\label{eqA12}
			D\equiv(1-\frac{\lambda _{BL} }{2\pi}),
                A\equiv\frac{\frac{M_{max}}{M_{\odot}}-b}{k},B\equiv\frac{\frac{R_{Mmax}}{km}-b}{k},
		\end{equation}
and $k$ and $b$ are the slope and intercept with the y-axis corresponding to each formula in Eq. (\ref{eq21}) and Eq. (\ref{eq22}), for example, when $\lambda _{BL}$ = 0.4, $D=1-({0.4}/{2\pi})$, $A=({{M_{max}}/{M_{\odot}}-0.0402})/{1506}$, $B=({{R_{Mmax}}/{km}-1.3229})/{903}$.

The Eq. (\ref{eqA9}) can also be written into the following form (we here only show the process of deducing from Eq. (\ref{eqA9}), for Eq. (\ref{eqA10}) it has a similar process),

\begin{equation}
           \label{eqA13}
   p_{rc}^{R-constraint}=DB^{2}(3{p^{2}_{rc}}+4p_{rc}\varepsilon_{c}+\varepsilon_{c}^{2}).
\end{equation}
Assuming that $p_{rc}$ can also be written in the polynomials of central energy density $\varepsilon_{c}$,

  \begin{equation}
			\label{eqA14}
p_{rc}^{R-constraint}=DB^{2}\varepsilon_{c}^{2}(1+aB^{2}\varepsilon_{c}+bB^{4}\varepsilon^{2}_{c}+cB^{6}\varepsilon^{3}_{c}+\cdots).
		\end{equation}
Putting Eq. (\ref{eqA14}) back into Eq. (\ref{eqA13}), and matching the coefficients, one have

 \begin{equation}
			\label{eqA15}
			a=4D,\; b=19D^{2},\; c=100D^{3},\cdots
		\end{equation}
Finally, the central EOS can be extracted from Eq. (\ref{eqA9}),

\begin{equation}
\label{eqA16}
    p_{rc}^{R-constraint}=DB^{2}\varepsilon_{c}^{2}(1+4DB^{2}\varepsilon_{c}+19D^{2}B^{4}\varepsilon_{c}^{2}+100D^{3}B^{6}\varepsilon_{c}^{3}+\cdots).
\end{equation}
For Eq. (\ref{eqA10}), it becomes
\begin{equation}
    \label{eqA17}
		p_{rc}^{M constraint} =DA^{\frac{2}{3}}\varepsilon_{c} ^{\frac{4}{3}}(1+4DA^{\frac{2}{3}}\varepsilon_{c} ^{\frac{1}{3}}+19D^{2}A^{\frac{4}{3}}\varepsilon_{c}^{\frac{2}{3}}+100D^{3}A^{\frac{6}{3}}\varepsilon_{c} ^{\frac{3}{3}}+\cdots).
\end{equation}

For the H model, the Eq. (\ref{eq7}) can be changed into
\begin{equation}
    \label{eqA18}
\widehat{p}_{t}=\widehat{p}_{r}-2\lambda _{H} \widehat{p}_{r} \frac{\widehat{m}}{r},
\end{equation}
combining with Eq. (\ref{eq8}) and Eq. (\ref{eq9}), it has $b_{1}=0, c_{1}=0, d_{1}=0, d_{2}=0, d_{3}=1/3$, and
\begin{equation}
			\label{eqA19}
			2b_{2}=-(\widehat{p}_{rc}+1)(\widehat{p}_{rc}+\frac{1}{3})+2(c_{2}-b_{2}),
		\end{equation}
\begin{equation}
			\label{eqA20}
			c_{2}-b_{2}=-\frac{2}{3}\lambda _{H} \widehat{p}_{rc}.
		\end{equation}
Thus
\begin{equation}
    \label{eqA21}
		b_{2}=-\frac{1}{6}[(\widehat{p}_{rc}+1)(3\widehat{p}_{rc}+1)+4\lambda _{H}\widehat{p}_{rc}], \; c_{2} =-\frac{1}{6}[(\widehat{p}_{rc}+1)(3\widehat{p}_{rc}+1)+8\lambda _{H}\widehat{p}_{rc}],
\end{equation}
and
\begin{equation}
           \label{eqA22}		
R\sim \frac{1}{\sqrt{\varepsilon_{c}}} (\frac{\widehat{p}_{rc}}{4\lambda _{H}\widehat{p}_{rc}+(\widehat{p}_{rc}+1)(3\widehat{p}_{rc}+1)})^{\frac{1}{2}}, \;
M\sim \frac{1}{\sqrt{\varepsilon_{c}}} (\frac{\widehat{p}_{rc}}{4\lambda _{H}\widehat{p}_{rc}+(\widehat{p}_{rc}+1)(3\widehat{p}_{rc}+1)} )^{\frac{3}{2}}.
\end{equation}
The central EOS becomes
\begin{eqnarray}
    \label{eqA23}
		p_{rc}^{M-constraint} =A^{\frac{2}{3}}\varepsilon_{c} ^{\frac{4}{3}}(1+DA^{\frac{2}{3}}\varepsilon_{c} ^{\frac{1}{3}}+(D^{2}+3)A^{\frac{4}{3}}\varepsilon_{c}^{\frac{2}{3}}+(D^{3}+9D)A^{\frac{6}{3}}\varepsilon_{c} ^{\frac{3}{3}}+\cdots ), \\
p_{rc}^{R-constraint}=B^{2}\varepsilon_{c}^{2}(1+DB^{2}\varepsilon_{c}+(D^{2}+3)B^{4}\varepsilon_{c}^{2}+(D^{3}+9D)B^{6}\varepsilon_{c}^{3}+\cdots),
\end{eqnarray}
where
\begin{equation}
			\label{eqA25}
			D\equiv4+4\lambda_{H}, \;
                A\equiv\frac{\frac{M_{max}}{M_{\odot}}-b}{k}, \; B\equiv\frac{\frac{R_{Mmax} }{km}-b}{k},
		\end{equation}
and the meanings of $k$ and $b$ are the same as in the BL model.

\section{Deravation of sound speed square}
For the BL model, one has
\begin{equation}
           \label{eqA26}		
M\sim \frac{1}{\sqrt{\varepsilon_{c}}} (\frac{\widehat{p}_{rc}}{4\lambda _{H}\widehat{p}_{rc}+(\widehat{p}_{rc}+1)(3\widehat{p}_{rc}+1)} )^{\frac{3}{2}},
\end{equation}
which is a function of central energy density $\varepsilon_{c}$. Taking derivative of $M$ with respect to $\varepsilon_{c}$ gives
\begin{equation}
           \label{eqA27}		
\frac{\mathrm{d}M}{\mathrm{d}\varepsilon_{c}}=\frac{M(\varepsilon_{c})}{2\varepsilon_{c}}[3(\frac{\varepsilon_{c}}{p_{rc}}\frac{\mathrm{d}p_{rc}}{\mathrm{d}\varepsilon_{c}}-1)\frac{1-3\widehat{p}_{rc}^{2}}{1+\widehat{p}_{rc}^{2}+4\widehat {p}_{rc}} -1].
\end{equation}
In addition, the derivative of $\widehat{R}$ with respect to $\widehat{p}_{rc}$ gives

\begin{equation}
           \label{eqA28}		
\frac{\mathrm{d}\widehat{R}}{\mathrm{d}\widehat{p}_{rc}}=\widehat{R}\frac{1-3\widehat{p}_{rc}^{2}}{2\widehat{p}_{rc}+8\widehat{p}_{rc}^{2}+6\widehat{p}_{rc}^{3}},
\end{equation}
and the derivative of $\widehat{p}_{rc}$ with respect to $\varepsilon_{c}$ gives
\begin{equation}
           \label{eqA29}		
\frac{\mathrm{d}\widehat{p}_{rc}}{\mathrm{d}\varepsilon_{c}}=\frac{\mathrm{d}\frac{{p}_{rc}}{\varepsilon_{c}}}{\mathrm{d}\varepsilon_{c}}=\frac{1}{\varepsilon_{c}}(\frac{\mathrm{d}p_{rc}}{\mathrm{d}\varepsilon_{c}}-\frac{p_{rc}}{\varepsilon_{c}} ).
\end{equation}
When ${\mathrm{d}M}/{\mathrm{d}\varepsilon_{c}}=0$ gives out the expression of sound speed square $s_{r,c}^{2}\equiv \mathrm{d}p_{rc}/\mathrm{d}\varepsilon_{c}$ as
\begin{equation}
           \label{eqA30}		
c_{r,s}^2 \equiv \frac{\mathrm{d}p_{rc}}{\mathrm{d} \varepsilon_{c}} =\widehat{p}_{rc}(\frac{1+3\widehat{p}_{rc}^2+4\widehat{p}_{rc}}{3(1-3\widehat{p}_{rc}^2)}+1),
\end{equation}
which is the same as in \cite{DTOV1,DTOV2}.

For the H model, it becomes

\begin{equation}
           \label{eqA31}		
c_{r,s}^2 \equiv \frac{\mathrm{d}p_{rc}}{\mathrm{d} \varepsilon_{c}} =\widehat{p}_{rc}(\frac{1+3\widehat{p}_{rc}^2+(4+4\lambda _{H} )\widehat{p}_{rc}}{3(1-3\widehat{p}_{rc}^2)}+1).
\end{equation}

\section{EOS extracted from PSR J0030+0451}
\begin{figure}[h]
\includegraphics[width=\columnwidth]{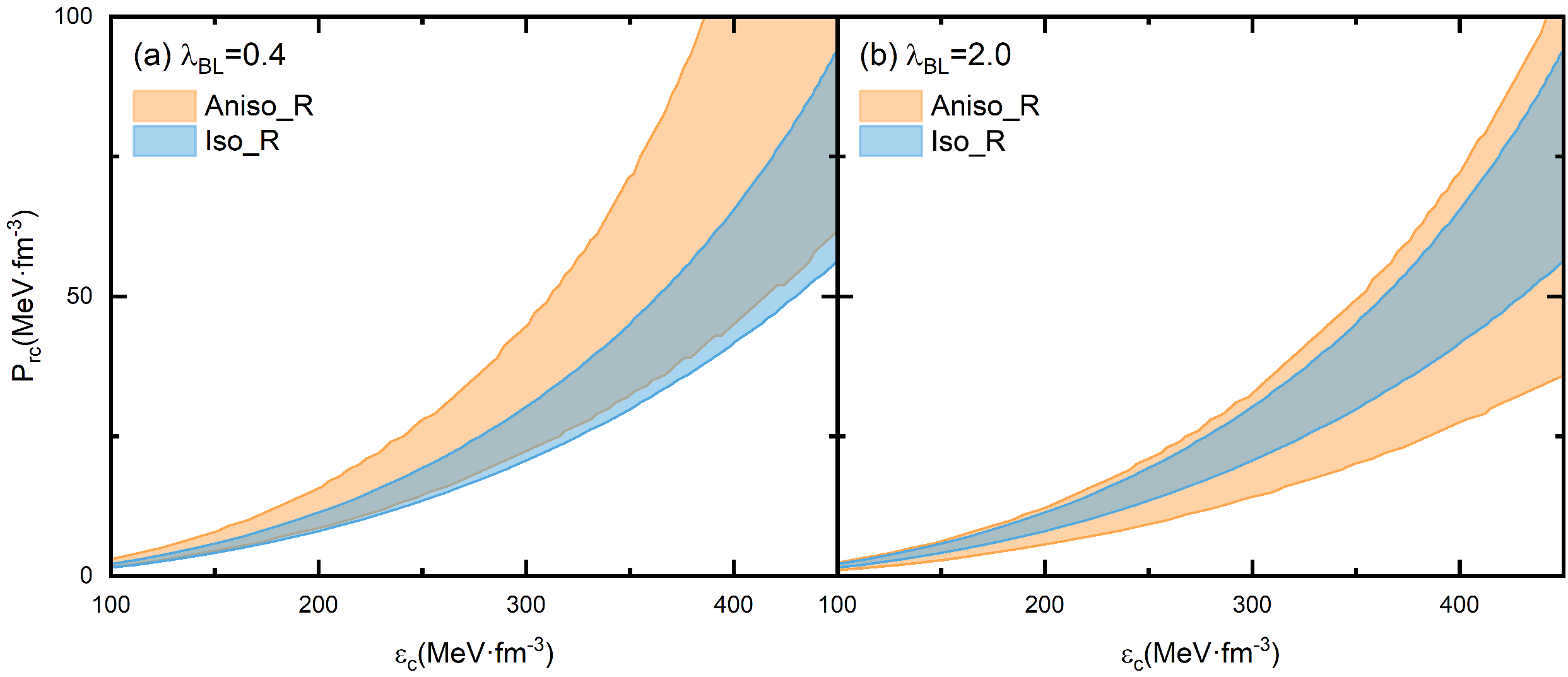}
\centering
			\caption{EOS extracted from the radius observation of PSR J0030+0451\cite{0451re}, for (a) $\lambda _{BL}$=0.4, (b) $\lambda _{BL}$=2.0.}
			\label{fig:8}
		\end{figure}
Besides the observation of PSR J0740+6620, the observation of a 1.4$M_{\odot}$ NS also gives a new constraint on NS radii, 11.96 $<$ $R_{1.4}$ $<$ 14.26 km \cite{04511}, and 11.52 $<$ $R_{1.4}$ $<$ 13.85 km \cite{04512}. Here we take the revised result 11.80 $<$ $R_{1.4}$ $<$ 13.10 km \cite{0451re}. Thus, as above, we can get the relation and the EOS constraint from the radius observation of PSR J0030+0451. However, for the maximum mass NS, different EOS predict a different maximum mass and a corresponding radius, but for the canonical neutron star, it predicts a different radius but with the same mass. Thus, we can only use the radius observation of PSR J0030+0451 now, with the future observation of the moment of inertia imposing another constraint on the central EOS of PSR J0030+0451.

For the BL model, we have  
\begin{eqnarray}
    \label{eqA32}
R_{\lambda_{BL}=0}^{1.4} &=& 0.671\times10^{3}\beta+2.783, \nonumber\\
		R_{\lambda_{BL}=0.4}^{1.4} &=& 0.354\times10^{3}\beta+6.765,  \nonumber\\
R_{\lambda_{BL}=1.2}^{1.4} &=& 0.311\times10^{3}\beta+7.353,  \\
        R_{\lambda_{BL}=2}^{1.4} &=& 0.255\times10^{3}\beta+8.232.  \nonumber
\end{eqnarray}
with r=0.948, 0.598, 0.517, 0.421 for $\lambda _{BL}$=0, 0.4, 1.2, 2.0.

For the H model, we have
\begin{eqnarray}
    \label{eqA33}
R_{\lambda_{H}=0}^{1.4} &=& 0.671\times10^{3}\beta+2.783, \nonumber\\
		R_{\lambda_{H}=0.4}^{1.4} &=& 0.423\times10^{3}\beta+6.136,  \nonumber\\
R_{\lambda_{H}=1.2}^{1.4} &=& 0.428\times10^{3}\beta+6.614,  \\
        R_{\lambda_{H}=2}^{1.4} &=& 0.435\times10^{3}\beta+7.075.  \nonumber
\end{eqnarray}
with r=0.948, 0.698, 0.752, 0.783 for $\lambda _{H}$=0, 0.4, 1.2, 2.0.

The constraint region from the BL model is showed in the Fig. \ref{fig:8}. As we have concluded, the pressure anisotropy also affect the extraction of central EOS of NSs.  However, compared to the result from PSR J0740+6620, the constraint from PSR J0030+0451 have a bigger uncertainty, which comes from the crust effect in the low mass NS.

\end{document}